# Influence of sample inhomogeneity on the linear elongational viscosities of two low density polyethylenes

*Teodor I. Burghelea, Helmut Münstedt*

*Institute of Polymer Materials, University Erlangen-Nürnberg, Martenstrasse 7, D-91058, Erlangen, Germany*

ABSTRACT: An experimental investigation of the impact of sample inhomogeneity on measurements of the linear elongational viscosity of two low-density polyethelenes is presented. A novel method of in-situ measurements of the diameter of samples under extension has recently been implemented to properly account for the sample non-uniformities during elongation. Two types of low density polyethylenes (LDPE's) have been investigated: Lupolen 1840 D and LPDE 1840 H. Whereas in the case of Lupolen 1840 H the Trouton relationship is verified, in the case of Lupolen 1840 D the deviations can be as large as 40%, depending on the magnitude of the initial sample non-uniformity and the experimental conditions. Based on real time visualization of the sample, these deviations are associated with an inhomogeneous deformation of the specimen. Differences in the homogeneity of deformation between the H and D samples are explained by significantly different maximal retardation times. The experimental investigation is complemented by a simplified theoretical estimation of the error induced by the sample inhomogeneity in the case of measurements of elongational viscosities in the linear range. A fair level of agreement is found with the experimentally measured error.

KEYWORDS: elongational viscosity, Münstedt rheometer, homogeneity of sample deformation

# I. Introduction

Pure extension is the dominant flow in many industrial processes and, therefore, accurate measurements of the extensional rheological properties of materials are very important. At a more fundamental level, such experiments play a crucial role in validating existing rheological models and suggesting new theoretical approaches. During the past three decades several experimental approaches to the extensional rheology of polymer melts have been proposed by Meissner [1], Münstedt [2], Sentmanat [3], Bach et al. [4] and variations of these methods by many others.

A comprehensive review of these different approaches to extensional rheology of polymer melts is not the object of the present work and can be found in [5] and more recently in [6].

It is well known that the homogeneity of deformation is crucial for reliably assessing the elongational properties of a material. Though recognized by many experimentalists this issue did not receive the proper attention, however. One reason for that is the difficulty to quantitatively measure the homogeneity of a sample, particularly, in the Sentmanat-type of rheometer. This situation is generally more promising to be tackled with for the Münstedt-type extensional rheometer as the sample is visible over its whole length but for an online measurement of the sample geometry some technical problems have to be overcome. With a new device developed at the *Institute of Polymer Materials* and thoroughly described in [7] it is possible now to get a reliable insight into the sample homogeneity during elongational measurements.

The online imaging of the sample in its various states of elongation is of big advantage as it avoids artifacts due to freezing in the specimen and measuring its diameters in the solid state after being removed from the rheometer.

Our criterion for the reliability of elongational measurements is the fact following from the linear theory of viscoelasticity that there exists the very simple relationship

$$\mu_0(t) = 3\eta_0(t) \quad (1)$$

with $\mu_0$ and $\eta_0$ being the linear elongational and the shear viscosity, respectively. This so-called Trouton relationship is valid for strain hardening materials only up to a maximum Hencky-strain of $\varepsilon_H \approx 1$. The Hencky strain has been defined as $\varepsilon_H = \ln\left(\dfrac{L(t)}{L_0}\right)$ where $L(t)$ and $L_0$ stand for the actual and initial length of the sample, respectively. As reliable elongational measurements are of interest up to $\varepsilon_H \approx 4$, assessments of the accuracy of experiments using eq. (1) are obviously of a somewhat limited value. But nevertheless methods to quantify the inhomogeneity of sample deformation and procedures using the data for corrections can be checked with respect to their applicability by investigations in the linear range of deformation.

We employ in this study an improved version of the technique we have presented in [7] to assess the validity of the Trouton relationship during measurements of the extensional viscosity in a linear range of deformation.

# II. Experimental section

**Experimental setup and techniques**

The experiments have been conducted with a Münstedt type extensional rheometer built in house which is sketched in Fig. 1(b). A detailed description of this device can be found elsewhere, [10]. The specimen **S** under investigation is clamped between the parallel plates $P_{1,2}$ of the rheometer and immersed in a silicone oil bath **C** to minimize gravity and buoyancy effects, Fig. 1(b).

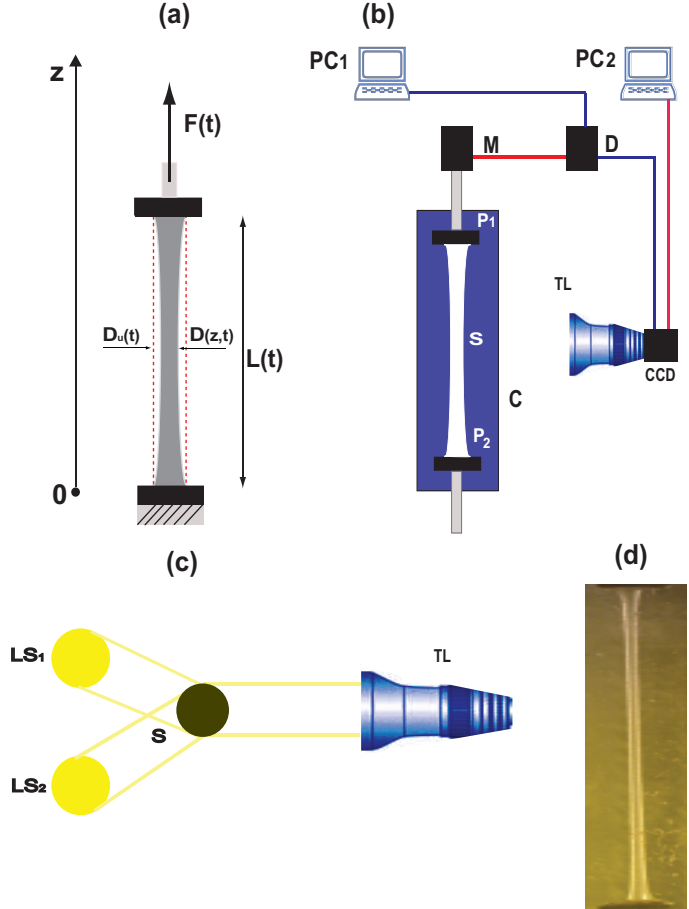

*Fig. 1: (a) Principle of the local measurements of the extensional viscosity. (b) Schematic view of the experimental arrangement: **C**- oil bath, $P_{1,2}$- top and bottom plates of the rheometer, **S**- the sample under investigation, **M**- AC servo motor, **D**- the control drive of the Münstedt rheometer, $PC_{1,2}$- personal computers, **TL**- telecentric lens, **CCD**- video camera. (c) Sample illumination and imaging: $LS_{1,2}$- linear light sources. (d) Example of a telecentric sample image corresponding to $\varepsilon_H = 2.7$ .*

While the bottom plate $P_2$ is stationary, the top plate $P_1$ is moved vertically by an AC-servo motor M, controlled by an analogue to digital converter installed on the computer $PC_1$. The sample is illuminated from behind by two linear light sources disposed as shown in the schematic top view presented in Fig. 1(c). The idea behind the backlight illumination arrangement is to obtain a maximum of brightness only on the edges of the sample and thus to allow accurate measurements of the sample diameter. A major challenge in properly imaging a considerably stretched filament comes from the high aspect ratio (height to width) of the corresponding field of view, which during our extensional experiments may be as large as 1:40. Thus, if a regular entocentric lens (with the entrance pupil located inside the lens) is used, both the spatial resolution and the level of geometrical distortion are unsatisfactory for high accuracy measurements of the sample diameter. Additionally, corresponding to large Hencky strains, both the illumination and the degree of focusing become uneven through the field of view. To circumvent these problems, we use in our study a high resolution telecentric lens with the entrance pupil located at infinity, (VisionMes 225/11/0.1, Carl Zeiss) which delivers images with very uniform brightness, free of distortions, perspective errors and edge position uncertainty. Sample images are acquired in real time using a high resolution (3000 by 1400 pixels) low noise camera (Pixelink from *Edmunds Optics*). The image acquisition is digitally synchronized with the rheometer. This allows us to directly compare the integral measurements of the transient elongational viscosity with the local ones and to correlate them both with the actual degree of uniformity of the sample.

## Materials and their rheological properties

Two materials have been used in this study: Lupolen 1840 D and Lupolen 1840 H. Several rheological properties of these materials are comparatively presented in Fig. 2.

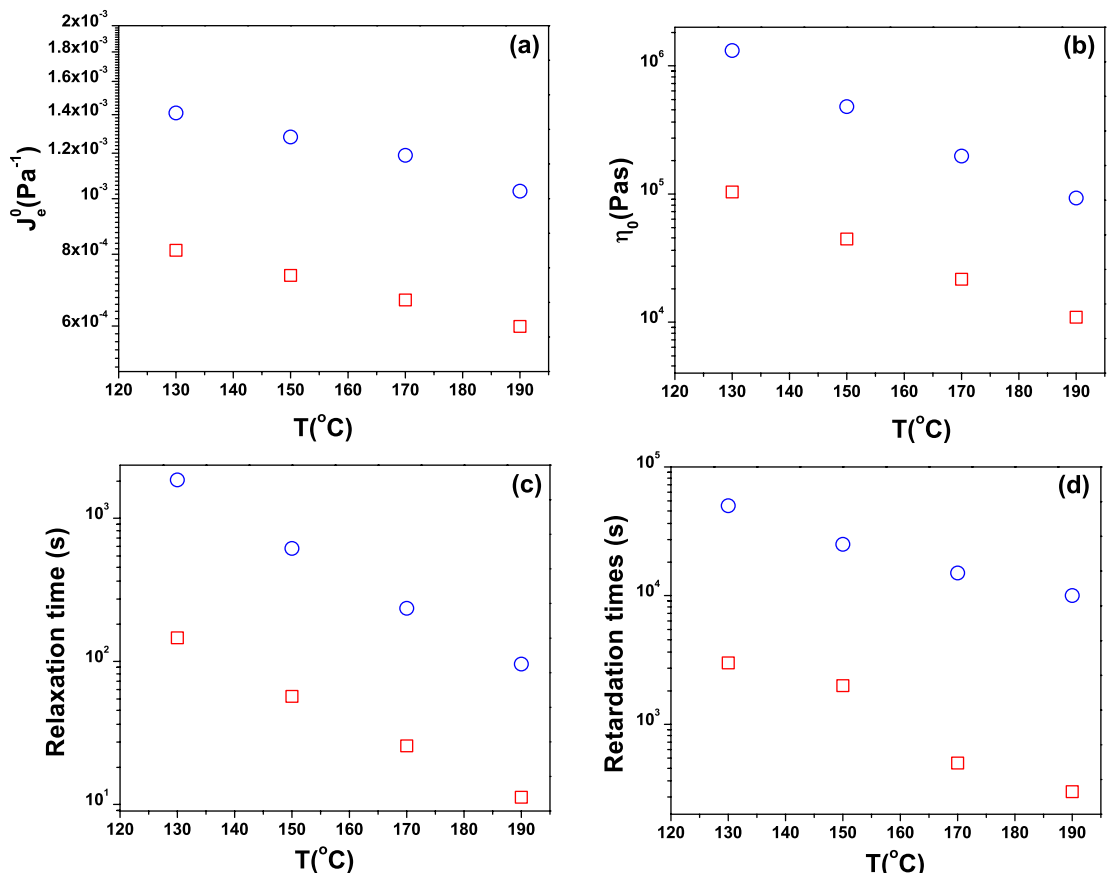

Fig. 2: Temperature dependence of rheological properties of the materials used: (a) Steady-state recoverable compliance (b) Zero shear viscosity (c) Relaxation time. (d) Retardation time.
The symbols are: circles -Lupolen 1840 D, squares- Lupolen 1840 H. Data is re-plotted from refs. [13] (Lupolen 1840 D) and [14] (Lupolen 1840 H).

At a given temperature, each of the relevant rheological parameters (zero shear viscosity, steady-state recoverable compliance, relaxation time and retardation times) is larger for the Lupolen 1840 D than for Lupolen 1840 H, mainly due to its higher molar mass. However, the largest differences (nearly two orders of magnitude) are visible in the retardation times obtained from creep experiments.

## III. Results

Fig. 3 shows measurements of the elongational viscosities as a function of time for the commercial LDPE Lupolen 1840 D at different elongation rates. The results are typical of this strain hardening polymer and frequently reported in the literature. If the envelope of the elongational viscosities which is composed of the curves up to $\varepsilon_H \approx 1$ is compared with the linear time-dependent shear viscosity $3\eta_0(t)$, however, a discrepancy of around 40 % is observed. The Trouton relationship is not verified indicating that the conditions of a reliable elongational experiment are not fulfilled.

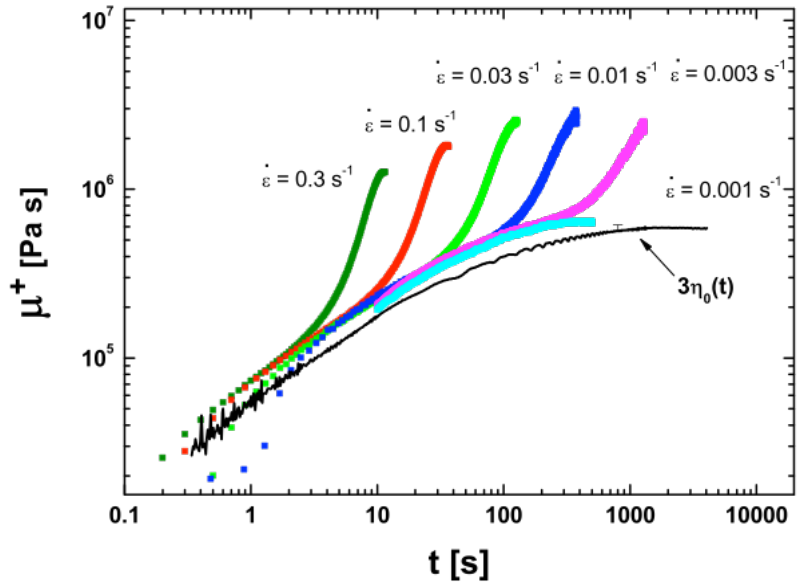

*Fig. 3: Extensional viscosities at different elongational rates (symbols) for Lupolen 1840 D at T=170 °C. The full line represents $3\eta_0(t)$ obtained by shear measurement. Data re-plotted from [13].*

'The results on Lupolen 1840 H follow the Trouton relationship with good accuracy, as Fig. 4 demonstrates.

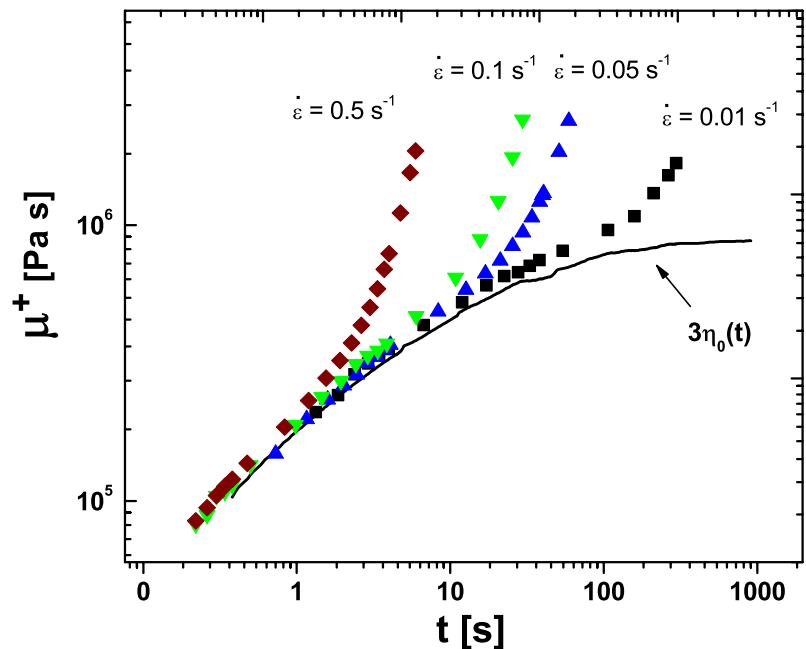

*Fig. 4: Extensional viscosities at different elongational rates (symbols) for Lupolen 1840 H at T=120 °C. The full line represents $3\eta_0(t)$ obtained from shear measurement. Data re-plotted from [14]. The data have been shifted from T=150 °C to T=120 °C using the time-temperature superposition principle and the shift factor is given in [15].*

These experimental findings which were presented by other authors in the literature, too, but not commented on at all (e.g. [9]) are presented in the following using results from the optical setup described in the experimental section.

Fig. 5 shows images of the specimens acquired at the same rate of deformation ($\dot{\varepsilon} = 0.01\ s^{-1}$) for Lupolen 1840 D at T=170 °C and Lupolen 1840 H at T=120 °C. The rate of deformation has been defined as $\dot{\varepsilon} = \frac{\varepsilon_H}{t}$. The experiments have been conducted at different temperatures in order to have comparable rheological properties for the two materials, Fig. 2.

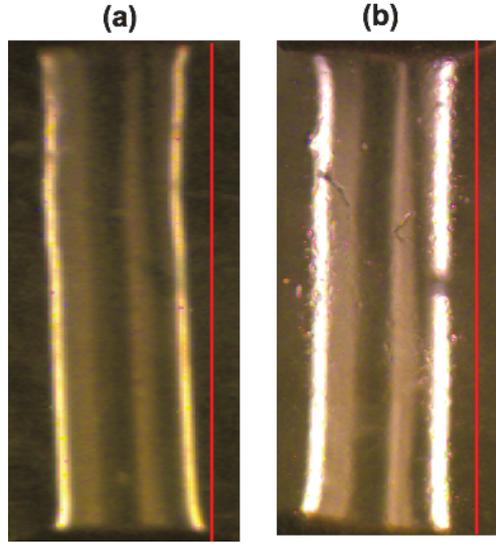

Fig. 5: Images of the samples elongated up to $\varepsilon_H = 0.5$ at $\dot{\varepsilon} = 0.01\ s^{-1}$ (a) Lupolen 1840 D, T=170 °C (b) Lupolen 1840 H, T=120 °C. The full lines added to each panel are guides for the eye and correspond to the initial location of the samples edge.

However, we point out that the retardation time of the Lupolen 1840 D at T=170 °C is still an order of magnitude larger than that of Lupolen 1840 H at T=120 °C.

Though apparently similar, there exists a major difference between the two images presented in Fig. 5: whereas for the Lupolen 1840 H sample the shape is cylindrical (note that the sample edges are parallel to the red guiding line), the shape of the Lupolen 1840 D sample develops a clear curvature. To quantify the magnitude of this effect the following non-dimensional inhomogeneity factor:

$$h(t) = \frac{D_{rms}(t)}{\overline{D}(t)} \qquad (2)$$

is defined with

$$D_{rms}(t) = \left\langle (D(z,t) - \overline{D}(t))^2 \right\rangle_z^{1/2}, \quad \overline{D}(t) = \left\langle D(z,t) \right\rangle_z \qquad (3)$$

In the equations above the brackets denote averages along the actual length of the sample, $L(t)$. Measurements of the inhomogeneity factors for both polymers investigated are presented in Fig. 6 up to a maximal deformation corresponding to $\varepsilon_H = 3$. Two distinct regimes may be observed in Fig. 6. At early stages of the deformation, the degree of inhomogeneity is constant and its value is probably related to the initial uniformity of the sample. A first interesting point is that a certain degree of sample non-uniformity exists in each of such extensional experiments and it is related to the initial sample positioning and adjustment of the top plate of the rheometer. Thus, this initial inhomogeneity hardly reproduces from experiment to experiment. As our materials have retardation times of the order of thousands of seconds the initial geometric inhomogeneity is likely to be "remembered" over a long period of the extensional test. A second point is that corresponding to a maximum in force, (see the inset in Fig. 6) the deformation process becomes increasingly inhomogeneous as clearly illustrated by the monotone increase of the degree of inhomogeneity, h. This corresponds to the onset of a non-uniform deformation of the sample, as predicted by the Considère criterion [11, 12]. As Fig. 7 demonstrates, the average sample diameter $\bar{D}$ is systematically larger than the ideal diameter $D_u$ corresponding to a uniform sample elongated at a constant rate, $\dot{\varepsilon}$. It has to be noted that although at large Hencky strains significant deviations from the idealized behavior (the full lines in Fig. 7) are observed for both materials, in the linear range ($\varepsilon_H \leq 1$) these deviations are more pronounced for the D species, Fig. 7. Thus, one should expect that the true stress calculated using $\bar{D}$ should be systematically smaller and thus closer to the time-dependent shear measurements than the integral measurement.

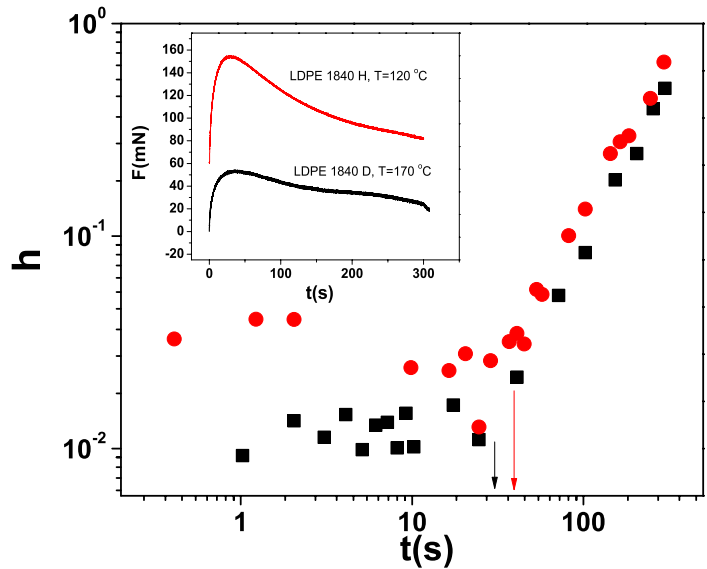

Fig. 6: Time dependence of the inhomogeneity factor, h. Circles: Lupolen1840 D, squares: Lupolen1840 H. The rate of extension was $\dot{\varepsilon} = 0.01\ s^{-1}$. The vertical arrows indicate the onset of an increasing sample non-uniformity. The force measurements are presented in the inset.

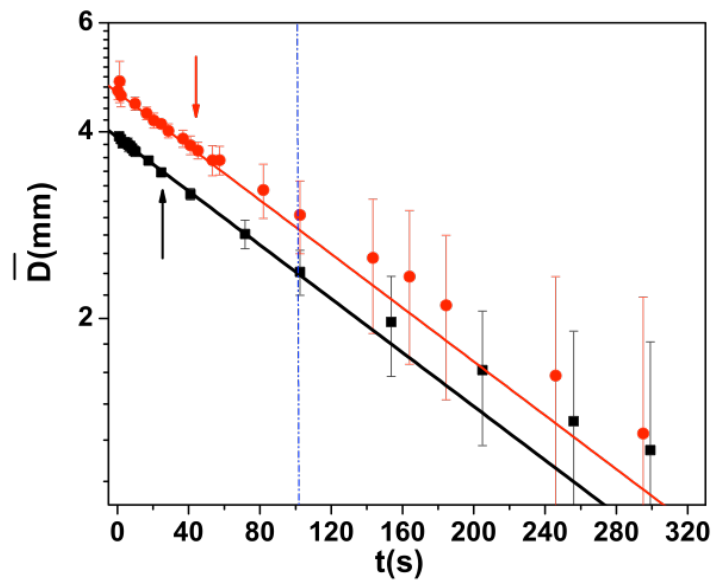

*Fig. 7: Transient average diameter of the sample during extensional experiments at $\dot{\varepsilon}=0.01\ s^{-1}$. The symbols are: circles -Lupolen 1840 D, T=170 °C, squares- Lupolen 1840 H, T=120 °C. The vertical arrows indicate the onset of the primary sample non-uniformity. The full lines are the diameters corresponding to a uniform $D_u$ deformation. The vertical dotted line indicates the extent of the linear range ($\varepsilon_H \leq 1$).*

A simple theoretical assessment of the impact of the geometric non-uniformity illustrated in Fig. 6,7 on the measurements of the transient elongational viscosity in the linear range is presented in the Appendix.

A more systematic discussion on the role of these deviations in the nonlinear range will be presented [16].

In Fig. 8 the integral viscosity measurements are displayed together with the local viscosities calculated from the measured force curves and the local average diameter from Fig. 7. As already stated above, for the Lupolen 1840 D, the loss of homogeneity translates into an underestimation of the true sample diameter and as a result, the true viscosity data taking the inhomogeneity into account fall systematically below the direct integral measurements but follow rather faithfully three times the shear data in the linear range, Fig. 8(a). A similar but significantly smaller effect is observed in the case of Lupolen 1840 H, Fig. 8(b).

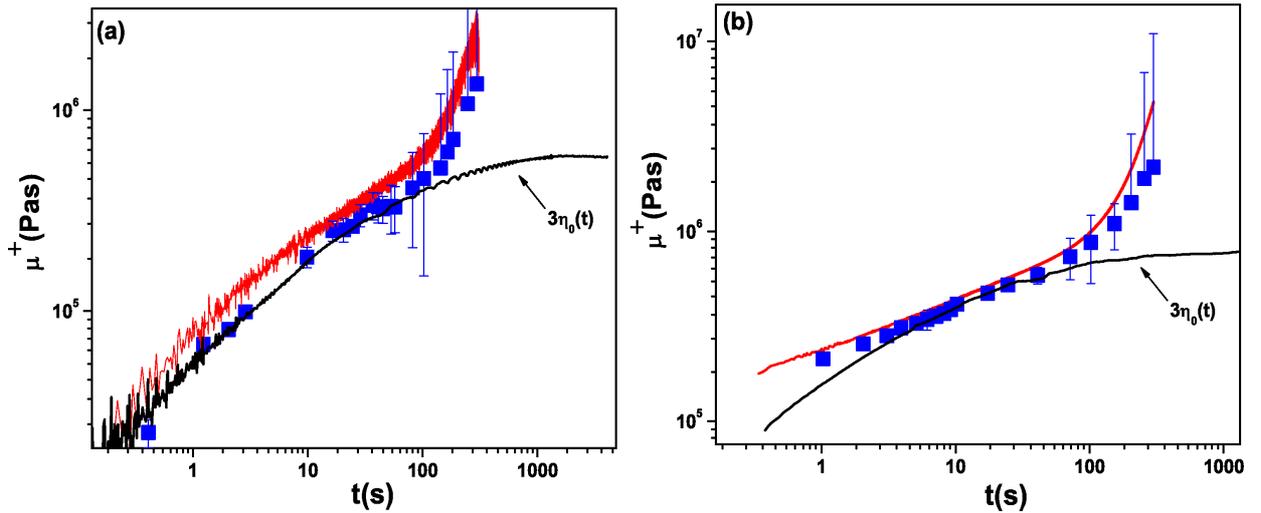

*Fig. 8: Comparison of the elongational viscosity $\mu^+(t)$ based on integral stress measurements (red curves), local stress measurements (symbols) at $\dot{\varepsilon}=0.01\ s^{-1}$ and three times the shear viscosity in the linear range (black curves): (a) Lupolen 1840 D, T=170°C (b) Lupolen 1840 H, T=120° C.*

In this case, except for the first few seconds (when the initial sample non-uniformities and alignment imperfections are still present) a reasonable agreement is found. The smaller effect in case of Lupolen 1840 H is easily explained by the better sample homogeneity, Fig. 6.

## IV. Discussion

After having clarified the reason for the discrepancy between elongational and shear data in the linear range by the non-uniformity of sample deformation the question arises on the relationship between the magnitude of this effect and the rheological properties of the material. A rather simplistic explanation would be the following. The rigid boundary conditions imposed by gluing the sample to the plates of the rheometer are always a source of geometrical non-uniformity regardless the molecular structure of the material simply because near the gluing points the materials deform less than in the bulk. Thus, a sample under extension at a certain elongation rate $\dot{\varepsilon}$ will always tend to develop a curvature in the proximity of the boundaries. If the deformation time scale, $\dot{\varepsilon}^{-1}$, is much shorter than the characteristic flow time of the material (interpreted here as the maximal retardation time), the sample will not have enough time to readjust itself to its natural cylindrical shape. As a consequence, the curved regions initially located near the plates will progressively extend towards the bulk of the sample leading to a systematic increase of the inhomogeneity factor, h. This hypothesis seems rather consistent with the data presented. As illustrated in Fig. 2 d the retardation times are more than an order of magnitude larger for the D species than for the H species. We also note that an initial degree of sample non-uniformity will always exist, regardless the nature of the material under investigation. This is clearly seen in Fig. 5, before a maximum in the tensile force is reached (at the left side of the arrows in the main plot). This effect is simply due to the inherent alignment imprecision during the loading of the sample and it is hardly reproducible in subsequent experiments. However, although such initial non-uniformities exist in principle for any material tested, their evolution with time is ultimately related to the rheological properties of the material, as discussed above.

# V. Appendix

A simplistic assessment of the impact of sample non-uniformity during measurements of the extensional viscosity is discussed. Let $D(z,t)$ be the real diameter of the sample measured at the position z along the sample (as indicated in Fig. 2(a)) and $D_u(t)$ the diameter of an ideal sample which preserves its cylindrical shape during the entire extension process. For a uniaxial deformation (which is the main assumption for most of the existing elongational rheometers), $D_u(t) = D_0 \exp(-\frac{\dot{\varepsilon}}{2}t)$ with $\dot{\varepsilon}$ the constant rate of elongation and $D_0$ the initial diameter of the sample. The integral elongational viscosity, $\mu^+(t)$, can be calculated by averaging the local stresses along the entire length of the specimen, $L(t)$:

$$\mu^+(t) = \frac{4F(t)}{\pi \dot{\varepsilon} L(t)} \int_0^{L(t)} \frac{1}{D^2(z,t)} dz \quad (4)$$

Here F(t) is the force acting on the sample. Taking into account that the elongational viscosity corresponding to an ideal (uniform) sample is given by $\mu_u^+(t) = \frac{4F(t)}{\pi \dot{\varepsilon} D_u^2(t)}$, it can easily be shown that the relative error induced by the sample non-uniformity (deviation from a cylindrical shape) is given by:

$$\frac{\mu^+(t) - \mu_u^+(t)}{\mu_u^+(t)} = -\frac{1}{L(t)} \int_0^{L(t)} \frac{4\delta(z,t)[D_u(t) + \delta(z,t)]}{(D_u(t) + 2\delta(z,t))^2} dz \quad (5)$$

where $\delta(z,t) = [D(z,t) - D_u(t)]/2$ quantifies the local deviation of the sample shape from the ideal cylindrical form. One can easily note from equation (5) that the absolute value of the relative viscosity error is bounded between 0 and 1 and thus, depending on the magnitude of $\delta(z,t)$, it may become quite significant. We also note that in the case that the idealized sample diameter $D_u(t)$ is locally smaller than the real sample diameter (which may happen if some parts of the sample deform slower than others due, for example, to the rigid boundary conditions near the clamping points), the true elongational viscosity $\mu^+(t)$ should fall below the idealized elongational viscosity $\mu_u^+(t)$ (which corresponds to a homogeneous deformation). This picture is quite consistent with the discrepancy being investigated here, where $\mu_0(t) = 3\eta_0(t)$ calculated from the shear data falls systematically below the measured extensional viscosity.

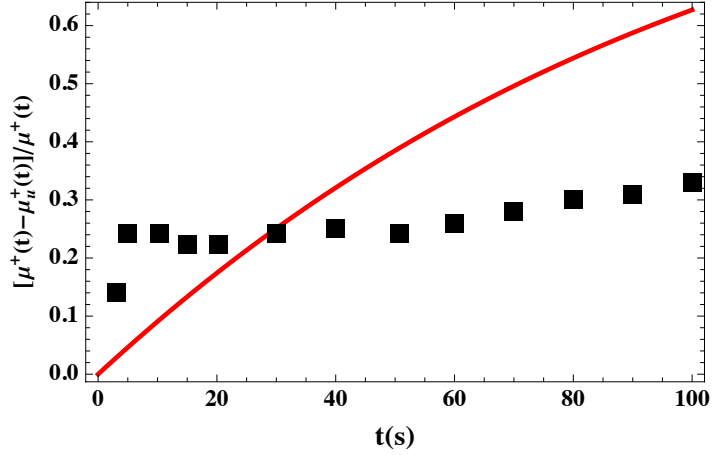

Fig. 9: Full line: Relative error of the transient elongational viscosity at $\dot{\varepsilon} = 0.01\ s^{-1}$ calculated using eq. (5) and a linear shape function, $\delta(z,t)$. Symbols: measured relative error using the data presented in Fig. 8 (a).

For a numerical calculation of the relative error introduced by the geometric inhomogeneity of the sample using equation (5), the transient shape function $\delta(z,t)$ needs to be obtained from a full numerical simulation of the elongation process. This task is nontrivial and probably unnecessary here, if one only needs to get a feeling of the magnitude of the error and further compare it with the experimental observation. For simplicity, we consider here a linear shape function defined by $\delta(z,t) = a\left|\left(z - \frac{L(t)}{2}\right)\right| + b$. The parameters $a, b$ are readily obtained imposing the rigid boundary conditions near the plates of the rheometer ($D[L(t),t] = D[0,t] = D_0$) and the incompressibility condition. In Fig. 9 we present both the calculated transient viscosity error (full line) and the measured error (the symbols) at $\dot{\varepsilon} = 0.01\ s^{-1}$, in a linear range, $\varepsilon_H \leq 1$. As the shape function $\delta(z,t)$ has been artificially prescribed, the numerical result presented in Fig. 9 should only be regarded as an indicator for the magnitude of the error in the viscosity measurements induced by the geometric non-uniformity of the sample. Although the onset of the primary sample inhomogeneity illustrated in Fig. 6,7 cannot be captured by this over simplified approach, the error in the viscosity measurement is rather consistent with the measured viscosity error in both magnitude and trend.

# VI. Acknowledgements

T. B. gratefully acknowledges the financial support from the German Research Foundation, grant $MU\ 1336/6-4$. We thank Mrs. Magdalena Papp for her assistance during some of the experiments reported in this study. One of us (T.B.) thanks Alfred Frey for the valuable technical advice, assistance with the Münstedt rheometer, and for the implementation of the digital trigger for the camera.